\documentstyle[twoside,fleqn,espcrc2,epsfig]{article}

\newcommand{\AmS}{{\protect\the\textfont2
  A\kern-.1667em\lower.5ex\hbox{M}\kern-.125emS}}

\newcommand{\be}{\begin{equation}}
\newcommand{\ee}{\end{equation}}
\newcommand{\ba}{\begin{eqnarray}}
\newcommand{\ea}{\end{eqnarray}}

\title{Properties of U(1) lattice gauge theory with monopole term 
        \thanks{This research was supported in part under DFG grant 
         Ke 250/13-1.}}

\author{G. Damm \address{Fachbereich Physik, Universit\"at Marburg,
		       D-35032 Marburg, Germany },
W.~Kerler \address{Institut f\"ur Physik, 
Humboldt-Universit\"at, D-10115 Berlin, Germany}}

\begin{document}

\begin{abstract}
In 4D compact U(1) lattice gauge theory with a monopole term 
added to the Wilson action we first reveal some properties of a third phase 
region at negative $\beta$. Then at some larger values of the monopole 
coupling $\lambda$ by a finite-size analysis we find values of the 
critical exponent $\nu$ close to, however, different from the Gaussian value.
\end{abstract}

\maketitle

\section{INTRODUCTION}

In recent higher-statistics studies of 4D compact U(1) lattice 
gauge theory with the Wilson action \cite{r97} and with this action 
supplemented by a double charge term \cite{cct97} for increasing lattice size 
the critical exponent $\nu$ has turned out to decrease towards $\frac{1}{4}$, 
the value characteristic of a first-order transition. 
Also stabilization of the latent heat has been observed. 
Thus, there are now rather strong indications that in those cases 
the transition is of first order. 
In contrast to this, for the action where a monopole term with coupling 
$\lambda$ is added, at $\lambda= 0.9$ the critical exponent has been found to
be characteristic of second order \cite{krw97}. 

In \cite{krw97} the location $\beta_{\mbox{\scriptsize cr}}$ of the
transition from the confinement phase to the Coulomb phase has been determined 
as a function of $\lambda$. It has been found that this transition line 
continues to negative $\beta$. On the other hand, in Refs.~\cite{hmp94,bhmp96} 
a further transition has been seen at $\beta=-1$ for $\lambda=0$ and at 
$\beta=-0.7$ for $\lambda=\infty$. Thus the question arises what 
happens at negative $\beta$ which is addressed in Sect.~2.

The energy distribution indicates that the second order region of the
confinement-Colomb transition starts at some 
finite $\lambda$ above $\lambda=0.7$ \cite{krw94}. An important question
is wether one has a region with universal critical properties there. To 
investigate this we have performed higher-statistics simulations 
also at $\lambda=1.1$ and at $\lambda=0.8$ and have evaluated the data for 
several variables by finite-size analyses. The respective results are presented 
in Sect.~3.

\section{PHASE REGIONS}

In \cite{krw97} the value $\beta_{\mbox{\scriptsize cr}}$ at the
transition from the confinement phase to the Coulomb phase has turned out 
to decrease with $\lambda$ and to get negative below $\lambda=1.2$ . 
The symmetry $\beta \rightarrow -\beta$, $U_{\Box} \rightarrow -U_{\Box}$
of the Wilson action at $\lambda=0$ gives rise to a transition at $\beta=-1$ 
in addition to the one at $\beta=1$ \cite{hmp94}. For $\lambda\ne 0$ the 
indicated symmetry is violated by the monopole term. At $\lambda=\infty$ only
the transition at negative $\beta$ persists and occurs at about 
$\beta=-0.7$ \cite{hmp94,bhmp96}. 

\begin{figure}[ht]
\vspace*{-5mm}
\psfig{figure=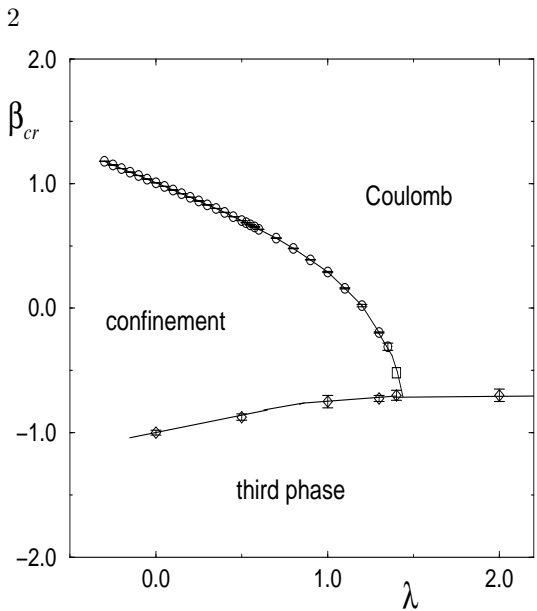,width=7cm,height=7.5cm}
\vspace*{-7mm}
\caption{Location of phase transition points $\beta_{\mbox{\scriptsize cr}}$
on $8^4$ lattice as function of $\lambda$ between confinement and Coulomb 
phases (circles from $C_{\mbox{\scriptsize max}}$, square from 
$P_{\mbox{\scriptsize net}}$) and to third phase (diamonds). }
\vspace*{-7mm}
\end{figure}

Here we have checked the occurrence of such transition at negative $\beta$ 
also at intermediate values of $\lambda$ determining the maximum of the 
specific heat $C_{\mbox{\scriptsize max}}$. It turns out that there is a 
transition line extending from $(\lambda,\beta)=(0,-1)$ to $(\infty,-0.7)$. 
Since from the properties we have observed so far we have no indication of a 
further subdivision of the region below this line, we consider it a third phase.
Figure 1 gives an overview of the phase regions as they are according to our 
present knowledge. It includes the line separating confinement and Coulomb 
phases obtained in \cite{krw97} and the data at negative $\beta$ found here. 

$P_{\mbox{\scriptsize net}}$, the probability to find an infinite network of 
monopole currents (where ``infinite'' on finite lattices is to be defined
in accordance with the boundary conditions), has turned out to provide an
unambiguous characterization of the confinement phase and the Coulomb phase
\cite{krw94}. For the periodic boundary conditions used here ``infinite'' 
means topologically nontrivial in all directions. Taking the values 
1 and 0 in the confinement and Coulomb phases, respectively, 
$P_{\mbox{\scriptsize net}}$ is very efficient to discriminate between those 
phases. In contrast to this in the third phase at fixed $(\lambda,\beta)$ 
we have observed values 0 as well as 1 for $P_{\mbox{\scriptsize net}}$. That 
$P_{\mbox{\scriptsize net}}$ provides no longer a criterion in the third phase
can be understood by noting that the monopole quantities are 
not invariant under the transformation $\beta \rightarrow -\beta$, 
$U_{\Box} \rightarrow -U_{\Box}$. 

As a characteristic feature of the third phase we have found that different 
states exist between which transitions in the simulations are strongly 
suppressed. We have observed this phenomenon at various negative $\beta$ in 
the $\lambda$ range from 0 to 2.5 . Typical examples of time histories of 
the average plaquette $\epsilon$ are given in Figure 2.

\begin{figure}[ht]
\psfig{figure=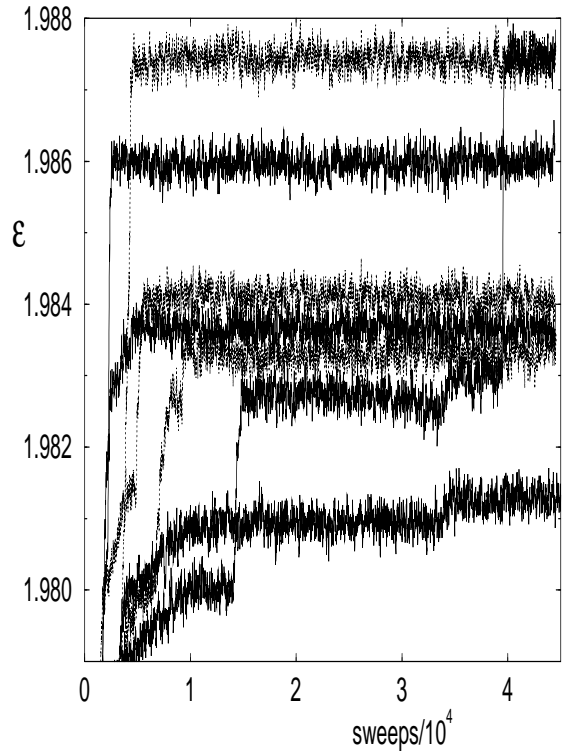,width=7cm,height=10cm}
\vspace*{-11mm}
\caption{Typical time histories of the average plaquette $\epsilon$
at negative $\beta$ obtained for seven different simulation runs,
shown for $(\lambda,\beta)=(2,-20)$ and $8^4$ lattice. }
\vspace*{-4mm}
\end{figure}

The values of $\epsilon$ increase as $\beta$ gets more negative. They are 
somewhat below 2 which indicates that at sufficiently negative $\beta$ the 
average value of $\cos \Theta$ gets close to $-1$. In view of the the symmetry 
$\beta \rightarrow -\beta$, $U_{\Box} \rightarrow -U_{\Box}$ 
of the Wilson action the correspondence of positive $\cos \Theta$
to positive $\beta$ and of negative $\cos \Theta$ to negative $\beta$
is conceivable.

The origin of the different states is not yet clear. Such states are similarly 
observed in spin glasses and frustrated systems, and also with 
spontaneous breaking of $Z(N)$ in finite-temperature SU(N) gauge theory.

\section{CRITICAL PROPERTIES}

\hspace{3mm}
At $\lambda=1.1$ and $\lambda=0.8$ for each lattize size considered Monte 
Carlo simulations have been performed at a number of $\beta$ values in the 
critical region. Multihistogram techniques  have been applied to 
evaluate the data and the errors have been estimated by Jackknife 
methods. In the finite-size analysis in addition to the specific heat and 
the Challa-Landau-Binder (CLB) cumulant complex zeros of 
the partition function, in particular the Fisher zero $z_0$ 
closest to the $\beta$ axis, have been used.

\begin{figure}[ht]
\vspace*{-2mm}
\epsfig{file=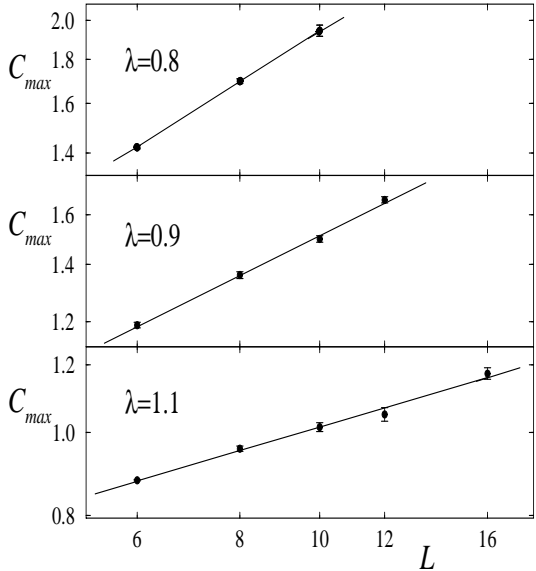,width=7cm,height=7.5cm}
\vspace*{-7mm}
\caption{ Maximum of specific heat $C_{\mbox{\scriptsize max}}$ as function
of lattice size $L$ for $\lambda=0.8$, 0.9 and 1.1 at transition
point $\beta_{\mbox{\scriptsize cr}}$ between
confinement and Coulomb phases.  }
\vspace*{-2mm}
\end{figure}

For $d=4$ the maximum of the specific heat is expected to behave as 
$C_{\mbox{\scriptsize max}} \sim L^4$
if the phase transition is of first order and as
$C_{\mbox{\scriptsize max}} \sim L^{\frac{\alpha}{\nu}}$
if it is of second order, where $\alpha$ is the critical exponent of the
specific heat and $\nu$ the critical exponent of the correlation length.

In Figure 3 we present the results for $C_{\mbox{\scriptsize max}}$ obtained 
on various lattices. They include data from simulations of the present 
investigation with $\lambda = 1.1$ and $\lambda = 0.8$ and ones from 
simulations of Ref.~\cite{krw97} with $\lambda = 0.9$ . 

The fits to the data in Figure 3 give values for $\frac{\alpha}{\nu}$ far
from 4, i.e. far from what would be expected for first order. Using these
values and the hyperscaling relation $\alpha = 2 - d\,\nu$ the values for 
$\nu$ listed in Table 1 are obtained. They are seen to be close to 
$\frac{1}{2}$. Thus in any case to conclude on second order appears quite safe. 

Similar results are obtained for the minimum of the CLB cumulant 
\ba
U_{\mbox{\scriptsize CLB}} &=& \frac{1}{3} (1-\frac{\langle E^4 \rangle}
  { \langle E^2 \rangle^2})
\label{fss5}
\ea
and for the imaginary part of the closest Fisher zero $z_0$. For these 
quantities finite-size scaling predicts the behaviors
\ba
  \mbox{Im}(z_0) &\sim& L^{-\frac{1}{\nu}}\quad ,\\
  U_{\mbox{\scriptsize CLB,min}} &\sim& L^{\frac{\alpha}{\nu}-4}
  \label{fss6} \quad .
\ea
The results of the respective fits are also listed in Table 1.

The values of $\nu$ obtained are seen to be close to the Gaussian value
$\frac{1}{2}$, however, different from it.
The observed increase of $\nu$ with $\lambda$ could indicate a 
nonuniversal behavior. Another possibility is that it is related to
finite-size effects. Then the increase should disappear on much larger
lattices. In that case the universal value of $\nu$ taken on the
infinite lattice could even be the Gaussian one.

\begin{table}[ht]
\caption{Critical exponents $\nu$ from Im($z_0$), $C_V$, 
$U_{\mbox{\scriptsize CLB}}$.}
\begin{tabular*}{75mm}{@{}l@{\extracolsep{\fill}}lllll}
%\begin{tabular}{lllll}
\hline
 $\lambda$ &  Im($z_0$)  & $C_V$  & $U_{\mbox{\scriptsize CLB}}$ \\
 \hline
   0.8   &  0.404(5)    &   0.433(2)  & 0.421(3) \\
   0.9   &               &   0.446(5)   &            \\
   1.1   &  0.421(8)     &   0.467(2)   & 0.455(2) \\
\hline
\end{tabular*}
\label{tab1}
\end{table}

\vspace*{2mm}
One of us (W.K.) wishes to thank
M.~M\"uller-Preussker and his group for their kind hospitality.

\end{document}